\documentclass[aps,pra,twocolumn,floatfix]{revtex4-1}

\usepackage{amsmath}
\usepackage{amssymb}
\usepackage{mathtools}
\usepackage{bbm}
\usepackage[caption=false]{subfig}
\usepackage[export]{adjustbox}
\usepackage{multirow}

\usepackage{setspace}
\usepackage{bm}
\usepackage{graphicx}
\numberwithin{equation}{section}
\usepackage{hyperref}

\makeatletter
\newcommand{\bs}[1]{\boldsymbol{#1}}
\renewcommand*\env@matrix[1][*\c@MaxMatrixCols c]{%
  \hskip -\arraycolsep
  \let\@ifnextchar\new@ifnextchar
  \array{#1}}
\makeatother

\usepackage{amsmath, amsthm, amssymb}

\newcommand{\be}{\begin{equation}}
\newcommand{\ee}{\end{equation}}
\newcommand{\bea}{\begin{eqnarray}}
\newcommand{\eea}{\end{eqnarray}}

\usepackage{rotating,wrapfig}
\usepackage{footnote}
\usepackage{tcolorbox}
\newtcolorbox{mymathbox}[1][]{colback=white, sharp corners, #1}
\usepackage{empheq}
\usepackage{upgreek}

\begin{document}

\title{Optical mode conversion in coupled Fabry-P\'erot resonators}
\author{Mark Stone$^1$}
\author{Aziza Suleymanzade$^1$}
\author{Lavanya Taneja$^1$}
\author{David Schuster$^1$}
\author{Jonathan Simon$^1$}
\affiliation{$^1$Department of Physics and James Franck Institute, University of Chicago, Chicago, IL}

\date{\today}

\begin{abstract}
Coherent control of the spatial properties of light is central to a wide variety of applications from high bandwidth quantum~\cite{vaziri2002experimental,krenn2014generation,nagali2009quantum} and classical~\cite{wang2012terabit,bozinovic2013terabit,richardson2013space} communication to high power fiber lasers~\cite{nicholson2012scaling}. Low-loss conversion amongst a complete and orthogonal set of modes is particularly important for robust mode-multiplexed communication~\cite{mirhosseini2015high,fontaine2019laguerre}. Here, we introduce tunable impedance mismatch~\cite{sweeney2019rsm} between coupled Fabry-P\'erot resonators as a powerful tool for manipulation of the spatial and temporal properties of optical fields. In the single-mode regime, frequency dependent impedance matching enables tunable finesse optical resonators, with potential applications in quantum science and sensing. Introducing the spatial dependence of the impedance mismatch as an additional ingredient enables coherent spatial mode conversion of optical photons at near-unity efficiency. We implement these ideas, experimentally demonstrating a NIR resonator whose finesse is tunable over a decade, and an optical mode converter with efficiency $>\!\!75\%$ for the first six Hermite-Gauss modes. We anticipate that this new perspective on coupled multimode resonators will have exciting applications in micro-~\cite{sabry2013silicon,xu2008silicon} and nano-~\cite{dai2013silicon,shen2014integrated,chen2015creating,burek2017fiber} photonics and computer-aided inverse design~\cite{piggott2015inverse}. In particular, combination with in-cavity electro-optics~\cite{huang2014proposal,soltani2017efficient} will open new horizons for real-time control of the spatio-spectral properties of lasers, resonators, and optical filters. 
\end{abstract}

\maketitle

\section{Introduction} \label{sec:intro}
High fidelity mode conversion and sorting are crucial tasks for quantum communication~\cite{vaziri2002experimental,krenn2014generation,mirhosseini2015high,nagali2009quantum}, as well as high-bandwidth mode-division multiplexed classical communication~\cite{wang2012terabit,bozinovic2013terabit,richardson2013space}. At the transmitting end of a communication network, mode conversion enables the encoding of information into the transverse spatial degrees of freedom of an optical field or fiber, thereby substantially increasing the bit-rate. At the receiving end, mode-sorting enables decoding of the previously encoded spatial information. While both mode conversion and sorting are fundamentally \emph{linear} in the electromagnetic field, they are technically challenging because the necessary linear transformations are not generically quadratic in the transverse spatial coordinates and as such cannot be directly implemented with standard optics like mirrors, lenses, and beam splitters.

At moderate efficiency, ``mode shaping'' can be achieved with a single phase plate~\cite{beijersbergen1994helical,demas2015free,bolduc2013exact} or digital micromirror device~\cite{zupancic2016ultra,mirhosseini2013rapid} that redirects a fraction of an incident optical field into a diffracted target mode; an additional spatial filter may be used to remove power in undesired modes~\cite{granata2010higher}. Near-unity efficiency requires implementing a unitary transformation of \emph{all} of the incident mode to the target mode. In the special case of Hermite-Gauss\,$\leftrightarrow$\,Laguerre-Gauss inter-conversion, this unitary transformation can be realized via a pair of astigmatic lenses~\cite{beijersbergen1993astigmatic}, since HG and LG modes are related to one another by only the relative phase of horizontal and vertical mode excitations. More general approaches to high fidelity mode-converting unitaries include numerically optimized nanostructured couplers between waveguides~\cite{lu2012objective,dai2015mode}; adiabatically varying coupling between macroscopic optical fibers~\cite{lai2007wavelength,leon2014mode}; conformal beam transformations implemented in two or more holographic phase gratings~\cite{PhysRevLett.105.153601,labroille2014efficient,huang2015mode,ruffato2018compact,fontaine2019laguerre}; meshes of Mach-Zehnder interferometers~\cite{miller2013self,ribeiro2016demonstration}; and long period fiber gratings~\cite{ramachandran2002bandwidth,li2015controllable}.

Implementing an arbitrary mode converter is formally equivalent to changing one quantum mechanical wave-function into another using only \emph{spatially local} potentials which cannot \emph{themselves} redistribute probability in space, but can impose phase gradients that result in such redistribution under the influence of a kinetic energy term. While lenses and mirrors can impart spatially varying phase profiles onto an incident optical field, it is the subsequent diffraction that must redistribute intensity; reshaping the mode via attenuation would irreversibly reduce the conversion efficiency. To our knowledge, all prior work fits within one of three paradigms: adiabatically varying the system Hamiltonian such that an input mode/initial eigenstate is smoothly converted into the desired output mode/final eigenstate (equivalent to coupled fibers); bang-bang unitaries that, in discrete steps separated by free-evolution/diffraction, convert between input and output modes (equivalent to cascaded diffraction gratings or long-period fiber gratings); or something in-between that implements a ``shortcut to adiabaticity''~\cite{del2013shortcuts}.

Here we present a new approach that breaks this paradigm and instead relies upon impedance mismatches between optical cavities to achieve near-unity efficiency mode conversion without nanophotonics or non-quadratic optics. Using only lenses and mirrors, we demonstrate conversion of an HG$_{00}$ mode into an arbitrary target HG$_{m0}$ mode by simply varying the length of a Fabry-P\'erot resonator over a few nanometers. The large propagation distances required for prior approaches are realized in our work by repeated round trips through the complex structure of the coupled cavities.

In Section~\ref{sec:tunable} we introduce the simpler problem of coupled, impedance-mismatched Fabry-P\'erot cavities in the \emph{single-mode} limit. Here the result is a cavity of tunable finesse $\mathcal{F}$. We then experimentally demonstrate such finesse tunability over a decade and characterize its properties in comparison with an S-matrix analysis. In Section~\ref{sec:converter} we consider the full problem of coupled, misaligned \emph{multimode} Fabry-P\'erot cavities, where the resulting behaviour corresponds to an optical mode converter. Implementing these ideas, we demonstrate optical conversion efficiency $>\!75\%$ for the first 6 Hermite-Gauss modes, limited by mirror loss and accidental mode-degeneracies. In Section~\ref{sec:outlook} we explore applications and outlook for these new tools.

\begin{figure}[h]
\includegraphics[width=.46\textwidth,valign=c]{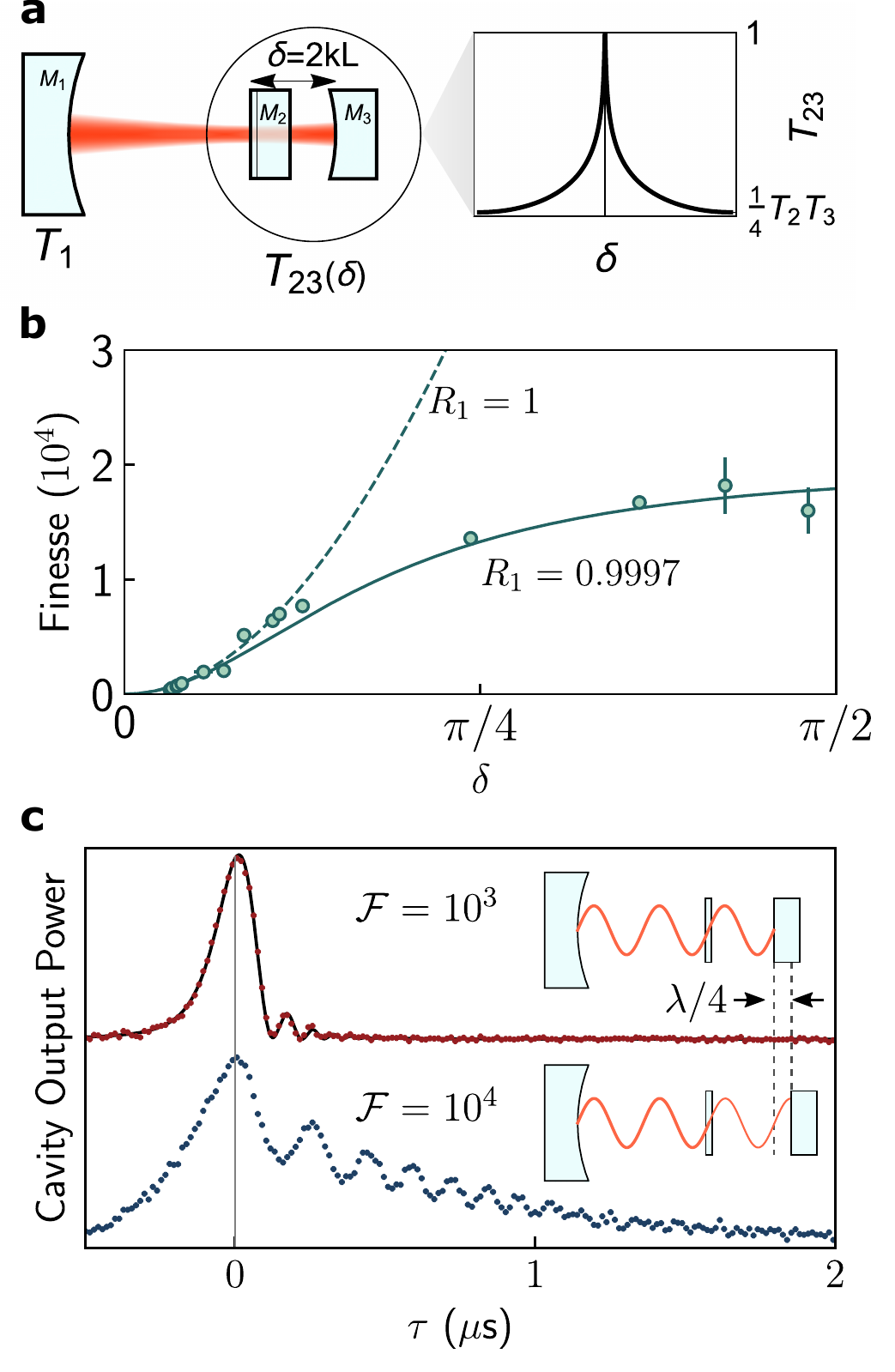}
\caption{\textbf{Tunable finesse optical cavity}. \textbf{a,} Schematic of two coupled single-mode cavities. Mirrors $M_2$ and $M_3$ act as a single ``effective mirror'' $M_{23}$ with frequency-dependent transmission $T_{23}(\delta)$, where $\delta\equiv 2kL$ is the round-trip propagation phase in $M_{23}$. Sub-$\lambda$ variations in the $M_2$-$M_3$ separation change their joint transmission, and thereby the finesse $\mathcal{F}$ of the composite $M_1$-$M_{23}$ cavity. $T_{23}$ varies from unity at resonance to $\frac{1}{4} T_2 T_3$ between resonances ($T_i$ is the power transmission of mirror $i$). \textbf{b,} Measured $\mathcal{F}$ of the $M_1$-$M_{23}$ cavity as a function of the round-trip optical phase $\delta$ in $M_{23}$, obtained from cavity ringdowns (for high $\mathcal{F}$) and transmission spectra (for lowest $\mathcal{F}$). The solid line is theory from measured mirror reflectances, limited by the reflectance ($R_1=0.9997(1)$) of $M_1$ ($R_1=1$ shown dashed). Error bars represent 1 s.d. of finesse. \textbf{c,} Typical low- and high-$\mathcal{F}$ ringdown measurements with representative exp-erfc fit~\cite{poirson1997analytical} (black line). Top inset: the cavities are mutually resonant (support an integer number of half wavelengths $\lambda/2$) for low $\mathcal{F}$. Bottom inset: displacing $M_3$ by a $\lambda/4$ shifts $M_{23}$ off resonance, reducing its transmission and achieving high $\mathcal{F}$.}
\label{fig:tunablefinesse}
\end{figure}

\section{Tunable Finesse Cavity} \label{sec:tunable}

We begin by analyzing two single-mode Fabry-P\'erot cavities coupled through a shared mirror, as shown in Fig.~\ref{fig:tunablefinesse}a. We will find that this arrangement acts as a tunable finesse cavity--- it traps light for a short duration (low finesse) or a long duration (high finesse). The two cavities have identical waists and share a mutual axis to avoid inter-mode coupling.

The total optical transmission of this arrangement can be calculated in the S-matrix formalism (see Appendix~\ref{SI:transfermatrixSM}) in terms of the lengths of the two cavities, the wavenumber $k$ of incident light, and the power reflection and transmission coefficients of the mirrors $M_1$, $M_2$, and $M_3$. A more intuitive understanding arises by observing that any single-mode scattering element is fully described by its (frequency $\delta\equiv 2kL$ dependent) reflection and transmission coefficients. It is thus valid to combine mirrors $M_2$ and $M_3$ with the propagation distance $L$ between them, into a single composite ``effective mirror'' $M_{23}$ with reflection and transmission coefficients $r_\textrm{23}(\delta)$, $t_\textrm{23}(\delta)$ (Fig.~\ref{fig:tunablefinesse}a, inset).

In this picture, what remains is the simple two-mirror ``primary'' cavity defined by the separation between $M_1$ and $M_{23}$. The total transmission of the primary cavity is thus precisely that of a simple two-mirror Fabry-P\'erot with a frequency-dependent reflection coefficient for one end-mirror. The finesse of the primary cavity can be computed according to~\cite{Siegman1986} as $\mathcal{F}\approx\frac{2\pi}{T_1+T_\textrm{23}+X_1+X_\textrm{23}}$ so long as the properties of $M_{23}$ remain $\sim$ constant across said resonance. Here $T_1$ and $T_\textrm{23}$ are power transmission coefficients and $X_1$, $X_\textrm{23}$ are power loss coefficients. As $T_\textrm{23}$ is tuned, the finesse $\mathcal{F}$ varies. Since $T_\textrm{23}$ is simply the transmission of the Fabry-P\'erot consisting of $M_2$ and $M_3$, it can range from unity to near zero as the length $L$ tunes the cavity from resonance to anti-resonance, thereby varying $\mathcal{F}$ from small to large values.

Harnessing these principles, we construct a tunable finesse cavity using mirrors with reflectances $R_1=0.9997(1)$ and $R_2=R_3=0.990(2)$ for 780 nm light. The finesse is tuned by varying $M_{23}$'s length with a piezoelectric actuator, and is then measured either spectroscopically or by cavity ringdown~\cite{poirson1997analytical} (Fig.~\ref{fig:tunablefinesse}c). The computed transmission of $M_{23}$, $T_{23}$ is shown in the black curve of Fig.~\ref{fig:tunablefinesse}a, varying from unity on resonance to $\frac{1}{4}T_2 T_3\approx 2.5\times 10^{-5}$ at maximum detuning. The measured finesse is shown in Fig.~\ref{fig:tunablefinesse}b, in close agreement with a parameter-free theory (solid curve). 

The finesse saturates at $1.7(2)\times 10^4$, limited by the reflectance of $M_1$, and is compared to theory for a perfect $M_1$ ($R_1=1$) in the dotted curve of Fig.~\ref{fig:tunablefinesse}b. From there, the next bound on finesse is set by the minimum transmission of the variable reflector, $\mathcal{F}_\textrm{max}=\frac{8\pi}{T_2 T_3}\approx 2.5\times 10^5$ for $R_2=R_3=0.99$. In practice, we anticipate an ultimate finesse limit set by scattering and absorption losses of the mirror coatings~\cite{hood2001characterization}, akin to a conventional Fabry-P\'erot cavity (see Appendix~\ref{SI:transfermatrixSM}). Since the $M_2$ substrate lies within the resonator, one might anticipate that absorption in the glass would strongly limit $\mathcal{F}$. However, in high-finesse configurations, very little power resides within $M_{23}$, so losses from the $M_2$ substrate and $M_3$ coating are strongly suppressed. A single-pass substrate loss of $1\%$ ($0.1\%)$ only limits $\mathcal{F}\leq 1\times 10^5$ ($2\times 10^5$), which improves further with higher $R_2,R_3$. Furthermore, fused-silica glasses can exhibit losses below 1 ppm/cm~\cite{Hild:06}, entirely obviating this limitation.

When the detuning between the cavities is smaller than the linewidth of the secondary cavity, the above picture breaks down, because (a) the $M_{23}$ transmission $T_{23}$ becomes strongly frequency dependent, or equivalently (b) there is an avoided crossing between the two coupled cavity modes. In practice, this means that the round-trip loss of the primary cavity cannot exceed $T_1+T_3$.

Imperfect mode matching leads to leakage of light out of the $M_1$-$M_{23}$ cavity through higher-order modes of $M_{23}$, potentially limiting the maximum achievable finesse. As with leakage through the lowest mode of $M_{23}$, this loss is suppressed as the modes are detuned from the primary cavity resonance. By making the $M_{23}$ cavity highly degenerate, it is possible for the fundamental mode of the primary cavity to be spectrally isolated from all modes of the $M_{23}$ cavity, thus avoiding accidental near-degeneracies. We choose $L$ to realize a half-confocal cavity with $\omega_{nlm}=\omega_{fsr}\left[n+\frac{1}{4}(l+m)\right]$, ensuring that the mode of the primary cavity is detuned by at least 1/8 of a free spectral range (FSR) from all modes of $M_{23}$. This detuning results in a transmission suppression of $\frac{2-\sqrt{2}}{4}\approx15\%$ relative to that at a  detuning of 1/2 the FSR: as long as the mode matching is better than $85\%$, the maximum finesse should not be significantly affected.

\begin{figure}[h]
\includegraphics[scale=.9,valign=c]{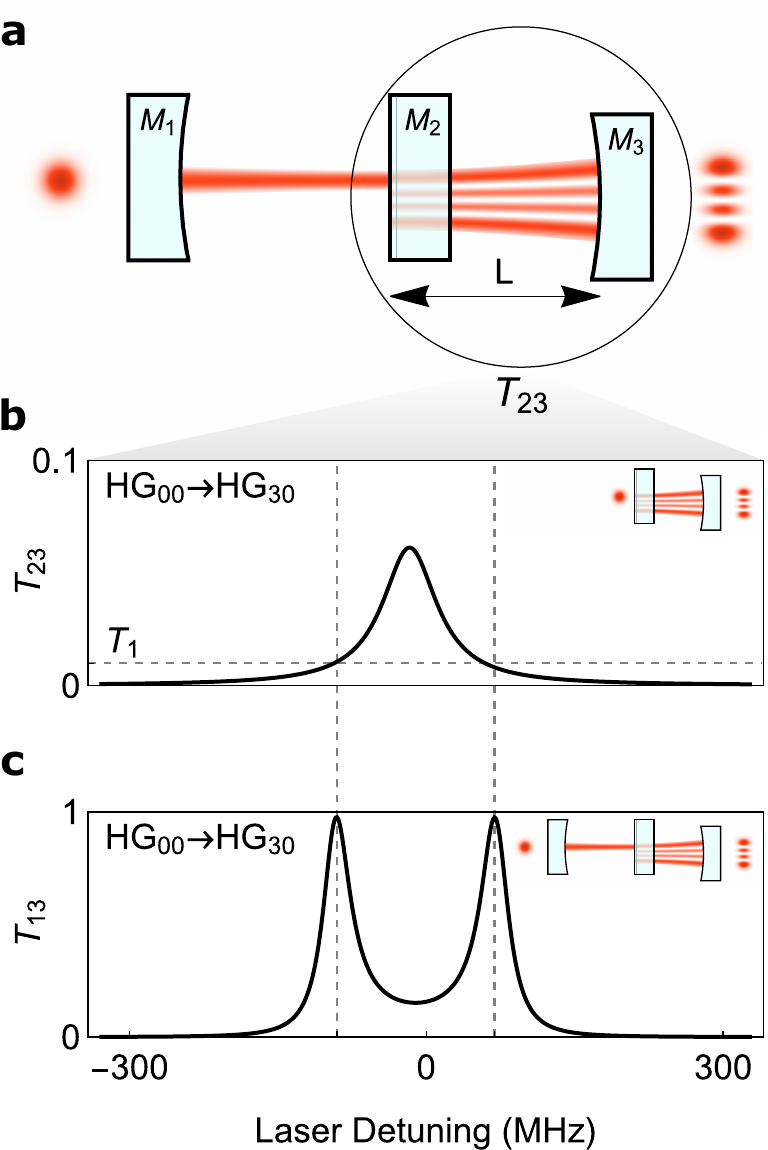}
\caption{\textbf{Principle of Optical Mode Conversion}. \textbf{a,} Two coupled Fabry-P\'erot resonators can act as an optical mode converter when a small transverse offset is introduced between their axes to couple their otherwise-orthogonal transverse modes. Mirrors $M_2$ and $M_3$ act as a single ``effective mirror'' $M_{23}$ with frequency- and mode- dependent transmission $T_{23}^{i\leftrightarrow j}$, for input/output modes HG$_{i/j,0}$. Near-unity efficiency $i\leftrightarrow j$ mode conversion through the full system $M_1$+$M_{23}$ is achieved when the input- and output- couplings to the composite cavity $M_1$/$M_{23}$ are equal, $T_1=T_{23}^{i\leftrightarrow j}$ (the ``impedance matching'' condition), and no light leaks out through other modes. \textbf{b,} Simulated transmission of the effective mirror $M_{23}$ (in the absence of $M_1$), with a translated HG$_{0,0}$ input generating an HG$_{3,0}$ output. The transmission $T_{23}^{0\leftrightarrow 3}$ is limited by the (translated) $0/3$ mode overlap of $\approx 6\%$, and the dashed horizontal line denotes $T_1$. The frequency dependence of the transmission guarantees that there are two frequencies where $T_1=T_{23}^{0\leftrightarrow 3}$ (dashed vertical lines), resulting in perfect mode conversion at these frequencies once mirror $M_1$ is introduced, as shown in \textbf{c}.}
\label{fig:modeconverter}
\end{figure}

\begin{figure*}[]
\includegraphics[width=\textwidth]{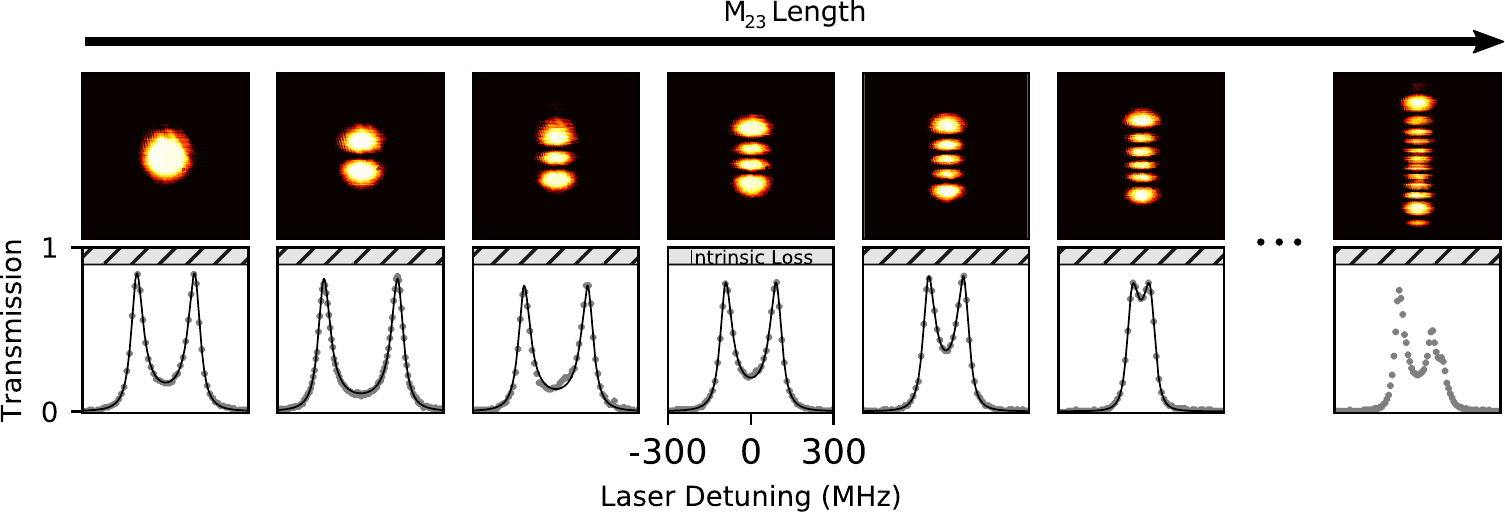}
\caption{\textbf{Demonstration of High-Efficiency Mode Conversion}. An input HG$_{00}$ mode may be coherently converted into any higher order HG$_{m0}$ mode by using two coupled, transversely offset Fabry-P\'erot cavities. The output spatial profile (top) and end-to-end conversion efficiency (bottom) are plotted for output modes HG$_{00}$...HG$_{50}$ and HG$_{10,0}$. As the length of output cavity $M_{23}$ is tuned with a piezoelectric actuator, its higher-order modes are individually brought near resonance with the drive laser. Each resonant mode of the $M_{23}$ cavity satisfies the impedance matching condition at two laser frequencies (Fig.~\ref{fig:modeconverter}), giving two peaks with near-unity efficiency mode conversion. In each panel the laser frequency is swept by $\pm 300$ MHz around the mutual resonance, demonstrating a mode-conversion bandwidth of $\sim\!50$ MHz. Mirror absorption and scattering limits the maximum conversion efficiency (hashed region). Optical power may be diverted into other accidentally degenerate modes, reducing conversion into the target mode and producing additional peaks and asymmetry in the transmission spectrum. Model fits (solid lines) are consistent with a transverse offset between cavities of $1.30(1)$ waists.}
\label{fig:allmodes}
\end{figure*}

\section{Mode Converter} \label{sec:converter}
Coupled optical cavities, as shown in Fig.~\ref{fig:modeconverter}a, enable near-unity efficiency mode conversion by a multimode generalization of single-mode impedance matching: at resonance, a two-mirror, single-mode cavity with equal in- and out- coupling $T_1=T_2$ transmits all light. The two coupled resonators explored in the prior section can be understood as one single-mode cavity with input coupling $T_1$ and (frequency-dependent) output coupling $T_{23}(\omega)$ of the composite mirror $M_{23}$. Unity transmission again occurs when in- and out- couplings are equal, $T_1=T_{23}(\omega)$. Because $T_{23}$ exhibits a resonance peak whose center frequency can be tuned by adjusting the length $L$ between $M_2$ and $M_3$, it is always possible to satisfy the impedance matching condition for a given drive frequency $\omega=\omega_d$.

In the absence of transverse mode coupling, an input $\textrm{HG}_m\equiv \textrm{HG}_{m0}$ mode produces an output HG$_m$ mode, and the single-mode analysis applies. Introducing a transverse offset between the coupled cavities breaks orthogonality between their higher-order modes and generates inter-mode couplings (Fig.~\ref{fig:modeconverter}a). In this case, the HG$_0$ mode of the primary cavity appears displaced on $M_{23}$, and thus has non-zero overlap with all modes of $M_{23}$. As such, $M_{23}$ now exhibits frequency- and \emph{mode}- dependent transmission $T_{23}^{i\leftrightarrow j}(\omega)=\left|\alpha_{ij}\right|^2 T_{23}^j(\omega)$, with input and output modes $i$ and $j$ having an overlap integral $\alpha_{ij}$. The transverse modes of $M_{23}$ each have their own transmission function $T_{23}^j(\omega)$, all with identical linewidths, but different resonant frequencies due to the round-trip Gouy phase of $M_{23}$~\cite{Siegman1986}. The simulated HG$_0\!\!\rightarrow\,\,$HG$_3$ transmission peak is shown in Fig.~\ref{fig:modeconverter}b.

We expect unity transmission to occur when $T_1=T_{23}^{i\leftrightarrow j}(\omega_d)$, where in-coupling occurs through the HG$_i$ mode at $M_1$, and out-coupling occurs through the HG$_j$ mode of $M_{23}$. The multimode S-Matrix calculation shown in Fig.~\ref{fig:modeconverter}c supports this intuition, showing nearly 100\% conversion efficiency. As the length of $M_{23}$, $L$, is tuned, its higher order modes individually approach resonance with the drive laser and primary cavity, satisfying the impedance matching condition at two drive-laser frequencies and thus permitting conversion of any input mode $i$ into any output so long as $\left|\alpha_{ij}\right|^2>T_1$. Indeed, for the theory in Fig.~\ref{fig:modeconverter}, the mode overlap between HG$_{0}$ and HG$_{3}$ is only $\sim6\%$, but near-unity conversion still occurs.

To demonstrate these principles, we construct a mode converter using mirrors with reflectances $R_1=R_3=0.965(5)$, $R_2=0.972(1)$ at 780 nm, whose performance is shown in Fig.~\ref{fig:allmodes}. In each panel, the laser frequency is scanned to satisfy the resonance condition. Between panels, the length $L$ of $M_{23}$ is varied with a piezoelectric actuator to bring the target mode to resonance with the $\textrm{HG}_{0}$ mode of the primary cavity. The transmission is monitored on a large-area photodiode to determine conversion efficiency, and on a CCD camera to ascertain mode shape. With the cavities transversely offset by $\sim$1 waist, $\textrm{HG}_{0}$ through $\textrm{HG}_{5}$ were generated with total conversion/transmission in excess of 75\%. To access higher order modes, the offset was increased to $\sim$2.5 waists and the piezo scanned as before, yielding conversion of modes up to $\textrm{HG}_{12}$; the $\textrm{HG}_{10}$ mode is shown with 75\% total transmission.

\section{Outlook} \label{sec:outlook}
We have presented a new paradigm for understanding coupled optical resonators, where one resonator acts as a frequency- and/or mode- dependent mirror for the other resonator. Harnessing this new perspective, we have demonstrated both a variable finesse optical resonator and an arbitrary spatial mode converter. By introducing an intracavity electro-optic modulator~\cite{huang2014proposal}, we anticipate rapid tunability of finesse and output mode, potentially enabling control of photon dynamics within a cavity lifetime.

In our approach, the mode conversion bandwidth is set by the cavity linewidth, and so can be increased by scaling down to micro-resonators. Working with small ROC fiber Fabry-P\'erots~\cite{hunger2010fiber} should enable bandwidths up to $\sim$10 GHz, and extending these ideas to nanophotonic platforms would allow further bandwidth gains~\cite{lu2011inverse,lu2012objective}.

The techniques introduced in this work can be employed to inter-convert between field profiles of \emph{any} physical system in which coupled resonators can be realized whose eigenmodes are the desired input- and output- field profiles. Coupling to a twisted optical resonator~\cite{schine2016synthetic} whose eigenmodes are Laguerre-Gauss (LG) would enable high-efficiency generation of optical orbital angular momentum states for optical communication~\cite{willner2015optical}. Similarly, the use of astigmagtic cavities would allow control over both mode indices of the HG$_{mn}$ output. Indeed, these concepts transcend even light: by coupling together phononic resonators with disparate mode structures, it should be possible to deterministically and efficiently reshape acoustic waves~\cite{whiteley2019spin,achilleos2017non}.

\section*{Methods}
The tunable finesse cavity and optical mode converter each consist of three low-loss dielectric mirrors supplied by LAYERTEC GmbH. All mirrors have fused silica substrates, rear-side anti-reflection coatings with reflectance $R<0.1\%$ at the operating wavelength of 780 nm, and front-side reflectances as described in the main text. Outer mirrors have concave surfaces while both middle mirrors are flat. Piezoelectric actuators are employed to vary the relative detuning between the two coupled cavities. A 780 nm distributed feedback laser (Eagleyard GmbH) provides light to test both setups. The beam passed through a 5m PM single-mode fiber to clean the spatial mode, yielding 15 mW of output power.

To measure the tunable finesse cavity, the laser frequency is swept across the cavity resonance by modulating the laser current, and the cavity transmission is measured on a fast photodiode (Thorlabs APD120A). At low finesse the \emph{frequency} width of the transmission peak reflects the cavity linewidth, with frequency scan calibrated against the transmission of 17 MHz sidebands induced by fast laser current modulation. At high finesse the cavity linewidth becomes smaller than the laser linewidth, and a different protocol must be employed.  The laser frequency is rapidly swept across the cavity line, and the linewidth is extracted from the ringdown waveform (see Fig.~\ref{fig:tunablefinesse}c and ref. ~\cite{poirson1997analytical}).

To measure the \emph{absolute} mode converter efficiency, light is picked off by two calibrated beam samplers before and after the converter and measured on large-area photodiodes (Thorlabs FDS100). The mode profile is measured on a CCD camera. 

\section*{Acknowledgements}
This work was supported primarily by AFOSR grant FA9550-18-1-0317. This work was also supported by the University of Chicago Materials Research Science and Engineering Center, which is funded by National Science Foundation under award number DMR-1420709. M.S. acknowledges support from the NSF GRFP.

\section*{Author Contributions}
M.S., A.S., D.S., and J.S. conceived the experiment. M.S. and L.T. performed the experiment. M.S. and J.S. developed the theoretical model. M.S. and J.S. drafted the manuscript. All authors contributed to the manuscript.

\section*{Author Information}
The authors declare no competing interests.

\bibliographystyle{naturemag}
\bibliography{library.bib}

\clearpage
\onecolumngrid
\appendix
\section{Single-Mode S-Matrix Approach for Coupled Fabry P\'erot Cavities} \label{SI:transfermatrixSM}
The behavior of a general linear coupled cavity system may be exactly analyzed with a scattering (S) matrix approach, so long as the paraxial and scalar field approximations are valid. In this section only a single spatial mode will be considered. The light field in a given transverse plane may then be described as an amplitude of a right- and a left-traveling wave, or a vector $\boldsymbol{\psi}=\left(\psi_r,\psi_l \right)^T$.

For a region of space containing paraxial optical elements between two transverse planes, there exists a mapping between the incoming waves on either side to the outgoing waves, called the scattering matrix. With the two sides labeled A and B, the scattering matrix $\bs{S}$ is defined by:

\begin{equation}
\label{eqn:smatdef}
\begin{bmatrix}
\boldsymbol{\psi}_{A,out}\\
\boldsymbol{\psi}_{B,out}\\
\end{bmatrix}
=
\begin{bmatrix}
S_{11} & S_{12}\\
S_{21} & S_{22}\\
\end{bmatrix}
\begin{bmatrix}
\boldsymbol{\psi}_{A,in}\\
\boldsymbol{\psi}_{B,in}\\
\end{bmatrix}
\end{equation}

The scattering matrices for simple optical elements like on-axis mirrors and regions of free propagation are well known. When multiple optical elements are placed in succession, the overall scattering matrix can be calculated by the transfer matrix approach, or equivalently by pairwise application of the cascaded scattering matrix formula:
\begin{equation}
\label{eqn:smatcascade}
\bs{S}^\textrm{tot}
=
\begin{bmatrix}
S_{11}^{1}+S_{12}^{1}S_{11}^{2}FS_{21}^{1} & S_{12}^{1}\left(1+S_{11}^{2}FS_{22}^{1}\right)S_{12}^{2}\\
S_{21}^{2}FS_{21}^{1} & S_{22}^{2}+S_{21}^{2}FS_{22}^{1}S_{12}^{2}
\end{bmatrix}
\end{equation}
with $F=\left(1-S_{22}^{1}S_{11}^{2}\right)^{-1}$.

This formula is sufficient to calculate the overall behavior of a single-mode paraxial system. Using the elementary scattering matrices for free propagation, $\bs{P}=\begin{bmatrix}0 & e^{i \phi/2}\\e^{i \phi/2} & 0\\ \end{bmatrix}$ and mirrors, $\bs{M}=\begin{bmatrix}r & it\\it & r \end{bmatrix}$, we obtain the scattering matrix for a Fabry-P\'erot resonator:
\begin{equation}
    \label{eqn:fpsmat}
    \bs{S}^{FP}=\begin{bmatrix}
         r_1 - \frac{e^{i \phi } t_1^2 r_2}{1-e^{i \phi } r_1 r_2} & -\frac{e^{\frac{i \phi }{2}} t_1 t_2}{1-e^{i \phi} r_1 r_2} \\
        -\frac{e^{\frac{i \phi }{2}} t_1 t_2}{1-e^{i \phi} r_1 r_2} & r_2 - \frac{e^{i \phi } t_2^2 r_1}{1-e^{i \phi } r_1 r_2} \\
        \end{bmatrix}\\
\end{equation}
where $\phi$ is the round-trip phase accrued in the cavity and $r_i,t_i$ are the field reflection and transmission coefficients. For two coupled Fabry-P\'erots the transmitted component is
\begin{equation}
\label{eqn:ccxmit}
S_{11}=\frac{-i e^{\frac{1}{2} i (\phi_1+\phi_2)} t_1 t_2 t_3}{1-e^{i \phi_1} r_1 r_2-e^{i \phi_2} r_2 r_3+e^{i\left( \phi_1+\phi_2 \right)} r_1 r_3 \left(r_2^2+t_2^2 \right)}
\end{equation}

An effective finesse for the primary cavity can be extracted by putting Equation~\ref{eqn:ccxmit} in the usual form of transmission through a Fabry-P\'erot, $E_t/E_i=-\frac{e^{\frac{i \phi_1}{2}} t_1 t_{23}}{1-g e^{i \phi_1}}$, with round-trip gain:
\begin{equation}
    g=r_1\frac{r_2-r_3 e^{i\phi_2} \left(r_2^2+t_2^2\right)}{1-r_2 r_3 e^{i \phi_2}}
\end{equation}
and $t_{23}$ the transmission of the $M_{23}$ cavity. Both of these numbers vary slowly with $\phi_2$ in high-finesse configurations.  Then the finesse is evaluated as~\cite{Siegman1986} $\mathcal{F}=\frac{\pi\sqrt{\left|g\right|}}{1-\left|g\right|}$.

The effect of mirror loss on the transmitted field can be easily calculated. Loss in the outer mirrors $M_1$ and $M_3$ simply reduces the transmitted power by a factor of $\frac{T_1}{1-R_1}\frac{T_3}{1-R_3}$, where $1-R_i$ is the power transmission of a lossless mirror with the same reflectance. To treat loss in $M_2$, we note that Equation \ref{eqn:ccxmit} is invariant under the substitution:
\begin{align*}
    r_2&\rightarrow r_2^\prime=\frac{r_2}{\beta}\\
    t_2&\rightarrow t_2^\prime=\frac{t_2}{\beta}\\
    r_1&\rightarrow r_1^\prime=\beta r_1\\
    r_3&\rightarrow r_3^\prime=\beta r_3\\
    t_1&\rightarrow t_1^\prime=\beta t_1\\
    \beta^2&=r_2^2+t_2^2=1-L_2
\end{align*}

Thus the transmitted field of a cavity with lossy $M_2$ is equivalent to a cavity with lossless $M_2$ and modified $M_1$ and $M_3$. This cavity has spectral properties set by $R_1^\prime,R_2^\prime,R_3^\prime$ and also a loss-induced amplitude reduction given by $\frac{T_1^\prime}{1-R_1^\prime}\frac{T_3^\prime}{1-R_3^\prime}=\beta \frac{T_1}{1-\beta^2 R_1}\frac{T_3}{1-\beta^2 R_3}\approx \frac{T_1}{T_1+L_1+L_2}\frac{T_3}{T_3+L_3+L_2}$ in the high reflectance limit.

Multielement scattering systems may also be treated as a signal flow graph and efficiently solved with Mason's gain formula~\cite{mason1953feedback}.

\begin{figure*}[]
\includegraphics[width=\textwidth]{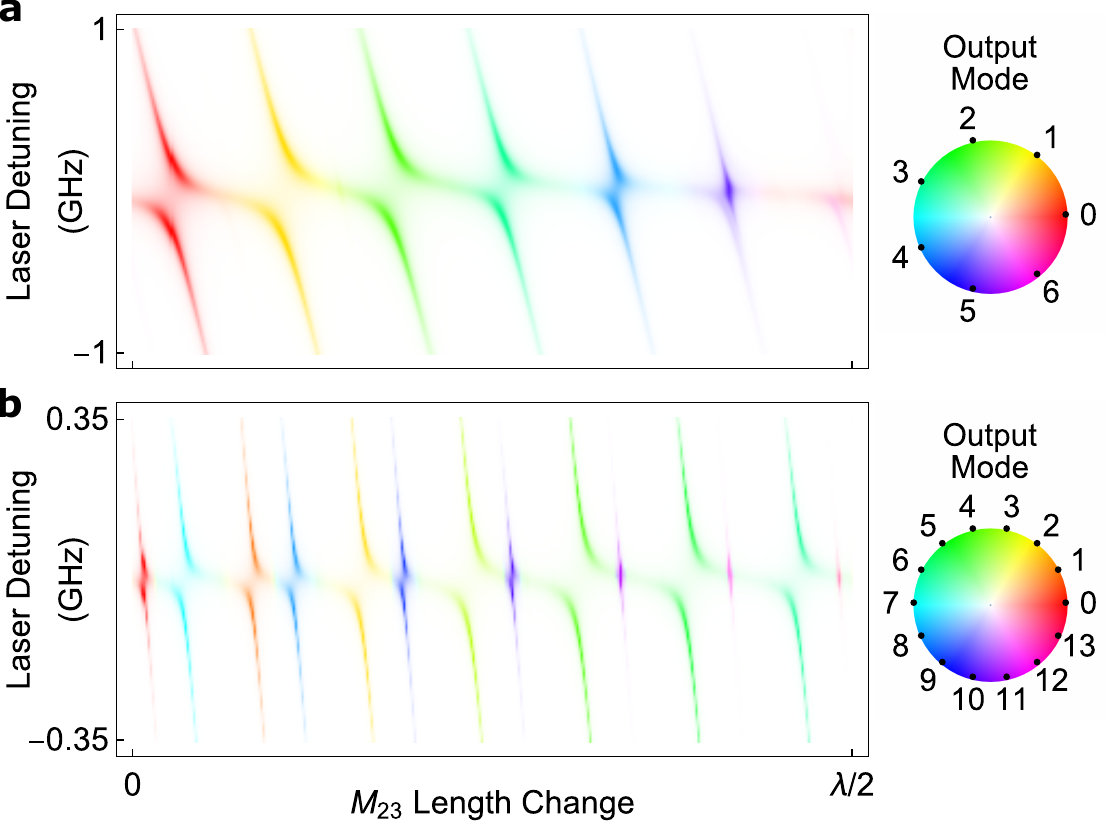}
\caption{\textbf{Simulated Spectrum of Coupled Multimode Optical Resonators}. \textbf{a,} Simulated transmission of the mode converter demonstrated in the main text, with $T_1=T_3=0.035$, $T_2=0.028$, and an input HG$_{00}$ beam. The output mode content (indicated by hue) varies as the length of the $M_{23}$ cavity is tuned, bringing different HG$_{m0}$ modes near resonance with the primary cavity. The output power (indicated by color saturation) reaches near unity at the impedance matched condition, as confirmed by the measured cross-sections in Fig.~\ref{fig:allmodes}, corresponding to perfect mode conversion. Successive mode orders display reduced peak splitting, reflecting a reduced coupling to the HG$_{00}$ mode of the primary cavity. After HG$_{50}$ the splitting is smaller than the cavity linewidth and impedance matching cannot be achieved, so higher-order modes disappear from the spectrum. The coupling coefficients are set by the transverse offset between cavities, here 1.3 mode waists. \textbf{b,} Increasing mirror reflectance ($T_1=T_2=T_3=0.01$) and transverse offset (2.1 mode waists) allows more modes to be impedance matched, and reduces leakage into accidentally near-degenerate modes. An input HG$_{00}$ can be coherently converted into HG$_{00}$--HG$_{13,0}$. Modes HG$_{70}$--HG$_{13,0}$ belong to the next lower axial mode group and appear interspersed amongst HG$_{00}$--HG$_{60}$.}
\label{fig:splitting}
\end{figure*}

\section{Multimode Scattering Matrix Approach for Coupled Fabry-P\'erot Cavities} \label{SI:transfermatrixMM}
A general paraxial system may be analyzed by the same scattering matrix approach with simple modifications. For simplicity this discussion will use a single transverse dimension, but the approach is easily extended to a full 2D transverse treatment. The light field in a given transverse plane may be decomposed into a basis of orthonormal Hermite-Gauss mode amplitudes for right- and left-traveling waves, described by a vector $\boldsymbol{\psi}=\left(\psi_{0,r},\psi_{1,r},\dots,\psi_{n,r},\psi_{0,l},\psi_{1,l},\dots,\psi_{n,l} \right)^T=\left(\boldsymbol{\psi}_r,\boldsymbol{\psi}_l \right)^T$. It is accurate to restrict to a finite number $n$ modes so long as the field distribution is bounded and nonsingular.

There is an infinite family of such Hermite-Gauss decompositions parameterized by the ``complex beam parameter'' $q$ and an axis around which the modes are centered. $q$ determines both the scaling (or waist) of the basis functions, and the degree of wavefront curvature.

Equations~\ref{eqn:smatdef} and~\ref{eqn:smatcascade} are both valid in the multimode case, but with the entries $S_{ij}$ understood to be block matrices of size $n \times n$. As long as the multimode scattering matrices of the individual optical elements are known, the overall scattering matrix may be calculated.

In the Hermite-Gauss basis, the scattering matrices of paraxial optical elements, such as free-space propagation and on-axis mirrors, have a simple form. Such elements do not produce mixing between modes, only overall rescaling and wavefront curvature, amounting to a change in the $q$ parameter~\cite{Siegman1986}. Put another way, the block elements $S_{ij}$ of the scattering matrix are diagonal, so long as it is understood that the fields on each port of the interface are expressed in Hermite-Gauss bases with the appropriate $q$ values. We could express the vector with reference to its basis $\boldsymbol{\psi}=\left(\boldsymbol{\psi}_{r,q_r},\boldsymbol{\psi}_{l,q_l} \right)^T$, but this will be left implicit in our equations. The required relation between $q$ values on each port of a paraxial element can be calculated using the ABCD matrix formalism, but it is not necessary for this discussion.

Inside an optical resonator, there exists a particular choice of $q$ which is transformed back into itself after each round trip~\cite{Siegman1986}. This is the most convenient choice, and it guarantees that every element of the resonator is described in a diagonal basis for most resonators, including all two-mirror resonators.

However, the mode converter consists of two optical resonators which have mismatched optical axes and/or waists, so no choice of optical axis and $q$ will yield diagonal forms for the scattering matrices of all elements. One solution is to describe the third mirror as an off-axis mirror whose scattering matrix has mode-mixing terms. Instead, we add an explicit change-of-basis matrix at the interface between the two resonators. This does not represent a physical optical element, but a mathematical transformation which allows the field on either side to be written in different bases. The matrix conveniently casts the field inside each resonator in terms of the eigenmodes of that resonator. The elements of the change-of-basis scattering matrix take the form
\begin{align*}
S_{12,mn}&=\langle\psi_m | \psi_n\rangle \\
S_{21}&=S_{12}^\dagger\equiv K^\dagger \\
S_{11}&=0 \\
S_{22}&=0
\end{align*}
where the overlap integral $\langle\psi_m | \psi_n\rangle$ between Hermite-Gauss modes with different optical axis and $q$ may be calculated numerically, or analytically using the method in Appendix~\ref{SI:overlap}.

The scattering matrices for paraxial elements with the correct $q$ are as follows. Mirrors act as $n$ copies of the form $\bs{M}$ from Appendix~\ref{SI:transfermatrixSM} on the individual modes. Propagation through free space gives a phase shift $P_{12,mn}=S_{21,mn}=\delta_{mn} e^{i \left[ k L+ \left( m+1 \right) \theta \right]}$, where $\theta$ is the well-known Gouy phase which may be calculated from $q$ and $L$.

The overall scattering matrix for mode-mismatched coupled cavities may be found from repeated application of Equation~\ref{eqn:smatcascade}. The left-to-right transmission is:
\begin{equation}
\bs{S}_{21}=-i t_1 t_2 t_3 e^{i \bs{\Phi_2}/2} \bs{K}\left[\bs{I}-r_1 r_2 e^{i\bs{\Phi_1}}-r_2 r_3 \bs{K}^\dagger e^{i\bs{\Phi_2}}\bs{K}+\left(r_2^2+t_2^2\right) r_1 r_3 e^{i\bs{\Phi_1}}\bs{K}^\dagger e^{i\bs{\Phi_2}} \bs{K} \right]^{-1} e^{i \bs{\Phi_1}/2}
\end{equation}
and the left-to-left reflection is:
\begin{equation}
\begin{split}
\bs{S}_{11}=r_1 \bs{I}-&t_1^2 e^{i\bs{\Phi_1}/2} \left[r_2 \bs{I}-\left( r_2^2+t_2^2\right) r_3 \bs{K}^\dagger e^{i\bs{\Phi_2}}\bs{K}\right]\times\\
&\left[\bs{I}-r_1 r_2 e^{i\bs{\Phi_1}}-r_2 r_3 \bs{K}^\dagger e^{i\bs{\Phi_2}}\bs{K}+\left(r_2^2+t_2^2\right) r_1 r_3 e^{i\bs{\Phi_1}}\bs{K}^\dagger e^{i\bs{\Phi_2}} \bs{K} \right]^{-1} e^{i \bs{\Phi_1}/2}
\end{split}
\end{equation}
where $\bs{\Phi}_i$ is the round trip propagation matrix for cavity $i$, including Gouy phases.

The mode converter of the main text is simulated using the S-matrix formalism in Fig.~\ref{fig:splitting}. Near-unity efficiency mode conversion is predicted when different modes of the two cavities are near resonant, as measured experimentally in Fig.~\ref{fig:allmodes}.

Multielement, multimode cavities may also be treated as noncommutative signal-flow graphs with matrix-valued weights and efficiently solved with Riegle's rule~\cite{riegle1972matrix}.

\section{Mode Purity}
\label{SI:modepurity}
Mode purity can be calculated exactly using the S-matrix formalism, or estimated from the mirror transmission coefficients. Mode purity is degraded due to imperfectly suppressed outcoupling through parasitic modes of the $M_{23}$ cavity. The mode in the primary cavity sees outcoupling through the target mode $T_{23}^{i\leftrightarrow j}$, which is set approximately equal to $T_1$ in the impedance matched condition. It also sees outcoupling through each unwanted mode $p$ equal to $T_{23}^{i\leftrightarrow p}=\left|\alpha_{ip}\right|^2 T_2 T_3 f_p(\omega)$, where $\alpha_{ip}$ is the mode overlap integral and $f_p(\omega)$ is a resonant enhancement factor, which is of order unity if the mode is moderately detuned. The ratio between power in parasitic mode $p$ and the target mode $j$ therefore scales as $\left|\alpha_{ip}\right|^2 \frac{T_2 T_3}{T_1} f_p(\omega)$. We note that the sum of all overlap integrals is bounded as $\sum_k \left|\alpha_{ik}\right|^2=1$. This estimate neglects the second-order effect of additional modes excited in the primary cavity.

\section{Overlap of Hermite-Gauss Modes} \label{SI:overlap}
To analyze coupled resonators which are not spatially mode-matched, it is useful to perform a change of basis between their eigenmodes. This transformation requires the overlap integrals between offset and/or rescaled Hermite-Gauss (HG) functions. These integrals may be calculated numerically, in which case it is useful to calculate the Hermite polynomials using a stable algorithm such as the recurrence relation~\cite{arfken1999mathematical} $H_{n+1}(x)=2x H_n(x)-2n H_{n-1}(x)$. They can also be calculated analytically using the method of generating functions. Here we work in a single transverse dimension for simplicity. The normalized Hermite-Gauss functions $\textrm{HG}_{n}$ with waist $w$ are given by:
\begin{equation}
    \textrm{HG}_{n}(x;w)=\sqrt{\frac{\sqrt{2/\pi}}{2^n n! w}} H_n \left(\frac{\sqrt{2} x}{w}\right) e^{-\frac{x^2}{w^2}}
\end{equation}
where $H_n (x)$ is the $n$th-order Hermite polynomial and the HG functions are taken to have no wavefront curvature (valid when the overlap is taken at the mode waist). The generating function for unnormalized HG functions is:
\begin{align}
    g_w (x,t)&=e^{\frac{2\sqrt{2} x}{w} t-t^2} e^{-\frac{x^2}{w^2}}\\
    &=\sum_{n=0}^\infty H_t \left(\frac{\sqrt{2} x}{w}\right) e^{-\frac{x^2}{w^2}} \frac{t^n}{n!}
\end{align}
The overlap integral between modes $m$ and $n$ of two HG bases is calculated by taking the integral of the product of their respective generating functions, picking off the correct series coefficients, and inserting normalization factors:
\begin{align}
    \int \textrm{HG}_{m,\lambda w}^* (x) \textrm{HG}_{n,w}(x+a w) dx &= \sqrt{\frac{2}{\pi 2^m m! 2^n n! \lambda w^2}} \left[ \frac{d^m}{du^m} \frac{d^n}{dt^n} \int g_{\lambda w}^* (x,u) g_w (x+a,t) dx \right]_{u,t=0} \\
    &= \sqrt{\frac{2}{2^m m! 2^n n!}} \sqrt{\frac{\lambda}{1+\lambda^2}} e^{-\frac{a^2}{1+\lambda^2}} \,\times\nonumber \\
    & \quad\left\{ \frac{d^m}{du^m} \frac{d^n}{dt^n} \textrm{Exp}\left[\frac{(1-\lambda^2)(t^2-u^2)+4tu\lambda+2\sqrt{2}a(u-\lambda t)}{1+\lambda^2} \right] \right\}_{u,t=0}
\end{align}

In this work we only require the overlap of modes with zero wavefront curvature. However, a similar derivation applies more generally, using the complex $q$-parameter formulation of the Hermite-Gauss functions.

\section{Two-Mode Coupled Mode Analysis}
\label{SI:twomode}
In Appendix~\ref{SI:cmt} we derive coupling constants for the phenomenological temporal coupled mode theory (TCMT) treatment of multimode coupled optical cavities. Here we use the results for a simple case with two transversely offset optical resonators coupled through a shared mirror, each supporting a single mode (which may have different transverse profiles). Assuming reciprocal media and neglecting loss, the scattering matrix takes the form~\cite{zhao2019connection,suh2004temporal}:
\begin{equation}
    S=-I-i\bs{D}\frac{1}{\left(\bs{\Omega}-i\bs{\Gamma}\right)-\omega}\bs{D}^T
\end{equation}
with
\begin{align}
    \bs{\Omega}&=\begin{bmatrix}
    -\frac{\delta}{2}&g\\
    g&\frac{\delta}{2} \end{bmatrix}\\
    \bs{D}&=\begin{bmatrix}
    \sqrt{\gamma_1}&0\\
    0&\sqrt{\gamma_2} \end{bmatrix}\\
    \bs{\Gamma}&=\bs{D}^\dagger \bs{D}=\begin{bmatrix}
    \frac{\gamma_1}{2}&0\\
    0&\frac{\gamma_2}{2} \end{bmatrix}\\
\end{align}
with $\delta$ the detuning between the modes. The coupling rate between the resonator mode in cavity $i$ and its corresponding output channel is $\sqrt{\gamma_i}=-\log R_i \nu_i$, where $\nu_i$ is the free spectral range and $R_i$ is the reflectance of the output mirror. The coupling rate between the two resonator modes is $g=\alpha \sqrt{-\log R_c \nu_1 \nu_2}$, where $\alpha$ is the overlap integral between the two modes and $R_c$ is the reflectance of the shared mirror.

Impedance matching occurs when the reflection coefficient vanishes. In the energy-conserving case, transmission reaches unity at this point, indicating full mode conversion. With matched cavity decay rates $\gamma_1=\gamma_2=\gamma$, this occurs at $\delta=0$, $\omega=\pm\sqrt{g^2-\left(\frac{\gamma}{2}\right)^2}$. Under these conditions, evaluation of the eigenmodes of the effective Hamiltonian, $\bs{\Omega}-i\bs{\Gamma}$, shows that equal stored energy resides in each cavity.

For $g<\frac{\gamma}{2}$ there is no real solution, but the minimum reflection occurs at $\delta=0,\omega=0$. Thus a solution with unit efficiency mode conversion exists whenever $\left|\alpha\right|^2>\frac{\gamma^2}{-4 \log R_c \nu_1 \nu_2}$.

For mismatched cavity decay rates $\gamma_1\neq\gamma_2$, impedance matching occurs at $\delta\neq0$, but there is still generally a solution for sufficiently large $\left|\alpha\right|$. In this case, the product of the stored energy and the decay constant of each cavity is equal.

\section{Multimode Coupled Mode Analysis}
\label{SI:cmt}
The S-matrix analysis of Appendix~\ref{SI:transfermatrixMM} relies only on the paraxial and scalar field approximations and is otherwise exact. Coupled optical cavities can also be analyzed using the temporal coupled mode theory (TCMT), a phenomenological model of open resonant optical systems. Although TCMT is not derived from first principles, it has been shown to agree well with rigorous analysis and provides useful intuition for the design of optical devices. TCMT is mathematically equivalent to the input-output formalism of damped quantum systems~\cite{gardiner1985input}. Here a full multimode theory for coupled optical resonators will be developed, while Appendix~\ref{SI:twomode} specializes to the two-mode limit to discuss impedance matching.

In this formalism an optical cavity is described by a set of $M$ cavity modes which are allowed to couple with each other and with $N$ ports, each containing an incoming and outgoing propagating channel. Assuming reciprocal media, the coupled mode equations are~\cite{zhao2019connection,suh2004temporal,sweeney2019rsm}    :
\begin{align}
    \label{eqn:ddta}\frac{d}{dt} \bs{a}&=-i\left(\bs{\Omega}-i\bs{\Gamma}\right)\bs{a}+\bs{D}^T \bs{s}_+\\
    \label{eqn:portcoupling}\bs{s}_-&=\bs{C}\bs{s}_++\bs{D}\bs{a}
\end{align}
where $\bs{a}$ is a state vector containing the $M$ amplitudes of the modes, normalized such that $\left|a_i\right|^2$ corresponds to the energy stored in the $i$th mode. $\bs{\Omega}$ and $\bs{\Gamma}$ are $M\times M$ Hermitian matrices, with $\bs{\Omega}$ representing the resonator mode frequencies and couplings and $\bs{\Gamma}$ representing decay processes. The resonances are coupled to the $N$ incoming channels $\bs{s}_+$ and outgoing channels $\bs{s}_-$ according to the coefficients in the $N\times M$ matrix $\bs{D}$. The channel amplitudes are normalized such that $\left|s_{+i}\right|^2$ ($\left|s_{-i}\right|^2$) is the power carried by the $i$th incoming (outgoing) channel. The $N\times N$ symmetric matrix $\bs{C}=\bs{C}^T$ represents direct coupling from input to output channels, including direct reflection and processes not included in the resonant modes $\bs{a}$.

Assuming harmonic time dependence for $\bs{a}$ then eliminating $\bs{a}$ from Equations~\ref{eqn:ddta}, \ref{eqn:portcoupling} gives the S-matrix $\bs{s}_-=\bs{S}\bs{s}_+$ as:
\begin{equation}
    \bs{S}=\bs{C}-i\bs{D}\frac{1}{\left(\bs{\Omega}-i\bs{\Gamma}\right)-\omega}\bs{D}^T
\end{equation}

For systems with no absorption loss, all decay comes from radiative coupling to propagating channels. When energy conservation and time-reversal symmetry hold, it can be shown that~\cite{zhao2019connection}:
\begin{align}
    \bs{\Gamma}&=\frac{\bs{D}^\dagger \bs{D}}{2}\\
    \label{eqn:cd}\bs{C}\bs{D}^*&=-\bs{D}\\
    \label{eqn:unitarity}\bs{C}^\dagger \bs{C}&=I
\end{align}
In what follows, we neglect loss so that these relationships hold.

All that remains is evaluation of the (system-dependent) coupling constants in $\bs{C}$, $\bs{D}$, and $\bs{\Omega}$. For two coupled optical resonators, the modes are enumerated as follows. All modes are labeled by their tranverse spatial mode index $t$. Channel modes have an additional port index yielding $s_{\pm,pt}$. Resonator modes have a cavity index and an axial mode index $z$ yielding $a_{ctz}$.

To define the direct coupling matrix $\bs{C}$ we note that any incoming power not coupled into the resonator is reflected into the same channel, so $\bs{C}$ is diagonal. Combined with Equation~\ref{eqn:unitarity}, this means each element of $\bs{C}$ is a phase factor with unit magnitude.  We take coupling to occur at the mirror surface, so that all transverse modes must experience the same reflection phase shift. This defines $\bs{C}$ up to a single arbitrary phase, which we choose so that $\bs{C}=-I$. 

The resonator-to-channel matrix $\bs{D}$ only couples modes with the same spatial mode index $t$. Resonator modes in a given cavity only couple to mirror(s) connected to that cavity. Therefore the element $D_{pt^\prime,ct z}=\xi_{ptz} \delta_{t,t^\prime} \sigma_{p,c}$, where we define $\sigma_{p,c}=1$ if port $p$ is connected to cavity $c$ and zero otherwise, and $\xi_{ptz}$ is a complex constant. The magnitude of $\xi_{ptz}$ is fixed by an energy conservation argument~\cite{haus1984waves}. We note that the energy of a single populated mode $a_{ctz}$ with no input decays as $\left|a_{ctz}(t)\right|^2=\left|a_{ctz}(0)\right|^2 e^{-\sum_p\gamma_p t}$, where the sum is over ports accessible from cavity $c$, $\gamma_p=-\nu_c \ln{R_p}$ is the decay rate into port $p$, $\nu_c$ is the free spectral range of cavity $c$, and $R_p$ is the reflectance of the mirror at port $p$. The power exiting is $\frac{d}{dt}\left|a_{ctz}(t)\right|^2=-\left(\sum_p\gamma_p\right)\left|a_{ctz}(t)\right|^2$. Therefore we ascribe a decay coefficient $\gamma_p$ to each port $p$ coupled to $a_{ctz}$. However, according to Equation~\ref{eqn:portcoupling} the power exiting into port $p$ is $\left|s_{-pt}\right|^2=\left|D_{pt,ctz}\right|^2\left|a_{ctz}(t)\right|^2$. Thus $\left|D_{pt,ctz}\right|^2=\gamma_p$ and $\left|\xi_{ptz}\right|=\sqrt{\gamma_p}$.

The phase of $\xi_{ptz}$ is constrained by Equation~\ref{eqn:cd} and our choice of $\bs{C}=-I$, yielding $\bs{D}^*=\bs{D}$, so all elements of $\bs{D}$ are real and defined up to a sign. Each resonant mode can have one arbitary sign in the coupling constant at one port. For all other ports accessible to that mode, the sign must be chosen consistently. This is important when multiple axial modes are included; adjacent axial modes have opposite parity, and incorrectly chosen signs will affect the interference between modes.

Finally we evaluate the closed-cavity Hamiltonian matrix $\bs{\Omega}$. The diagonal elements are just the (real) mode frequencies set by the free spectral range and transverse mode spacings. The off-diagonal elements represent coupling rates between resonant modes. We evaluate these with a similar energy conservation argument as used for $\bs{D}$~\cite{haus1984waves}. Coupling occurs between the modes of two cavities separated by a mirror of reflectance $R_c$. The circulating power in mode $a_{c^\prime t^\prime z^\prime}$ of cavity $c^\prime$ excites a mode $a_{ctz}$ of cavity $c$. According to Equation~\ref{eqn:ddta}, the coupling contributes to $\frac{d}{dt}a_{ctz}$ a term $\Omega_{ctz,c^\prime t^\prime z^\prime} a_{c^\prime t^\prime z^\prime}$. This can be compared to excitation of a mode by a propagating channel, which contributes to $\frac{d}{dt}a_{ctz}$ a term $D_{pt,ctz} s_{+pt}$, where the incident power is $P=\left|s_{+pt}\right|^2$ and we have already determined the magnitude $\left|D_{pt,ctz}\right|=\sqrt{-\nu_c \ln{R}}$. In the present case the incident power due to mode $a_{c^\prime t^\prime z^\prime}$ is $P=\left|a_{c^\prime t^\prime z^\prime}\right|^2 \nu_{c^\prime}\left|\alpha_{t^\prime,t}\right|^2$, where the overlap integral $\alpha_{t^\prime,t}$ restricts to that portion of the incident mode which is spatially mode-matched. Comparing these two cases, we must have $\left|\Omega_{ctz,c^\prime t^\prime z^\prime}\right|=\sqrt{-\nu_c \nu_{c^\prime} \ln{R_c}}\left|\alpha_{t^\prime,t}\right|$. The phase of the coupling coefficients must be chosen with similar concern as the elements of $\bs{D}$, taking into account the opposite parity of adjacent axial modes.

Although there exists an exact coupled-mode description of single-mode resonators~\cite{lang1988exact} which could be extended to the multimode case, we do not pursue that here, as the S-matrix description of Appendix~\ref{SI:transfermatrixMM} provides exact results, and the simpler coupled-mode theory is quite accurate and useful for intuition.

\newpage
\clearpage
\newpage

\onecolumngrid

\end{document}